# Growth and Photoelectrochemical Study of Germanium Sulphoselenide GeS$_{0.25}$Se$_{0.75}$ (I$_2$) Crystals

Dr. Sandip Unadkat[⊥], Dr. Love Trivedi[†], Aastha Anish Patel[*]

[⊥]Birla Vishvakarma Mahavidyalaya Engineering College, Vallabh Vidyanagar

[†*] Department of Physics, JPIS, Jaipur, Rajasthan, India

**Abstract**

In the present investigation, the author has employed a Chemical Vapour Transport (CVT) technique to grow the crystals of GeS$_{0.25}$Se$_{75}$ using iodine as a transporting agent. The grown crystals were then characterized for a Photoelectrochemical(PEC) study to find out solar parameters e.g. Fill Factor (FF), Open Circuit Voltage ($V_{oc}$), Short Circuit Current ($I_{sc}$), and Efficiency (n). The found results have been thoroughly described and implications have been discussed.

## Introduction

Our planet earth receives an abundant amount of solar energy each year. Which is mainly consisting of ultraviolet (approx.. 3%), visible region(approx.. 44%), near-infrared(approx. 52%), and rest is far-infrared regions [1].

Solar energy conversion is a very developed area of microelectronics, but there are still some problems that require proper attention. One of the major problems which most scientists and researchers are facing is the storage of electrical energy. This problem can possibly be solved with the help of electrochemical solar cells. [2,3]. Photoelectrochemical cells (PECs) extract electrical energy from light, including sunlight. Each cell consists of one or two semiconducting photoelectrodes and also auxiliary metal and reference electrodes immersed in an electrolyte. [4, 5].

Among the many semiconducting materials investigated as electrodes in the PEC solar cells, the metal chalcogenides have attracted considerable attention because they have energy band gaps well suited to solar energy conversion, a high absorption coefficient in the visible range and extremely good stability when in contact with various aqueous and non-aqueous electrolytes. The basic crystal unit of these compounds is a structure of three planes, a metal (T) is a sandwich between two chalcogens (X) planes. Within a layer of X-T-X, a strong covalent bond exists whereas the layers are held together by much weaker Van der walls interactions. In addition to their use in PEC solar cells, they are also used as electrode materials for cheap batteries offering a rather high density [6,7].

**Experimental:**

**Growth:**

In the present investigations, the high-quality crystals of $GeS_{0.25}Se_{0.75}$ ($I_2$) were grown by the chemical vapor transport (CVT) technique using iodine as a transporting agent. The main requirement of this technique is the precise setting of the temperature gradients between two zones to enhance the transport material in vapor form. For this purpose, a dual-zone horizontal furnace having the required dimensions was used which is shown in Fig. 1. The furnace was constructed at the University Science and Instrumentation Centre (USIC), Sardar Patel University by using a special sillimanite threaded tube closed at one end, 450 mm in length, 70 mm outer diameter, 56 mm inner diameter with a threaded pitch of 3 mm, imported form from Koppers Fabriken Feuerfester, Germany.

High-quality quartz ampoules were used for the growth experiment having dimensions of 24 cm in length, 2.4 cm outer diameter, and 2.2 cm inner diameter. The ampoule was first washed with detergent powder and boiled water after then with a hot mixture of concentrated $HNO_3$ and $H_2SO_4$ taken in an equal proportion, followed by washing of the ampoule with distilled water. After this ampoule was filled with concentrated HF and heated until HF evaporated so that the inner surface of the ampoule becomes rough which enhances the nucleation process during the growth. In the end, this ampoule was washed at least 8 to 10 times with double distilled water to remove any residue of these chemicals inside the ampoule. A cleaned ampoule was kept in a constant temperature furnace at 1000C for nearly 24 hours to make it moisture free.

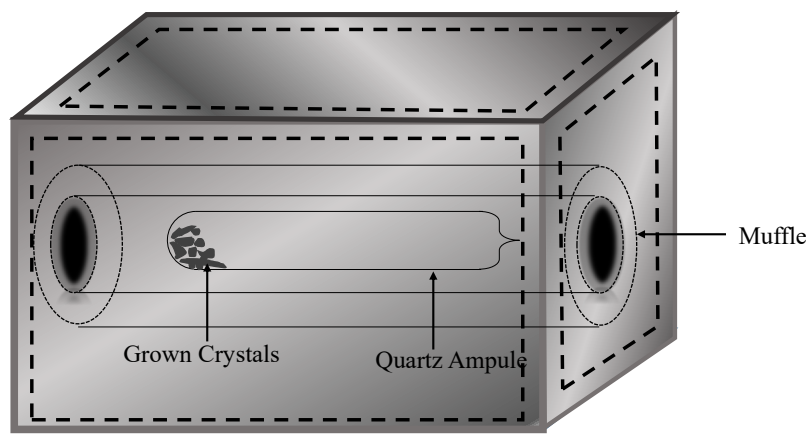

*Figure 1:* *The dual-zone horizontal furnace with co-axially*

For compound preparation, the cleaned ampoule was filled with a stoichiometric proportion of high purity Ge (99.999%), S (99.99%), and Se (99.99 %) of about 10 grams for growth then the ampoule was sealed under the pressure of $10^{-5}$ Torr. The sealed ampoule was kept in a dual-zone horizontal furnace. The temperatures of both the zones were slowly but gradually raised to the desired temperature and maintained that temperature for 3 days after the furnace was cooled off at room temperature. The ampoule was broken and shaken well with help of agate mortar to prepare the fine powder of this compound. For the growth of crystals, this compound was filled again into another chemically cleaned ampoule and repeated the procedure mentioned above. The temperature of both zones of the furnace raised slowly from room temperature to require temperature at the rate of 40K/hr and kept the temperature steady for seven days. After the growth period, the temperature of both zones cooled at the rate of 20K/hr to room temperature. The grown crystals were carefully retrieved from the ampule after breaking it. The photograph of grown crystals of $GeS_{0.25}Se_{0.75}$ ($I_2$) is shown in Fig. 2. The growth condition for the grown crystal is mentioned in Table 1

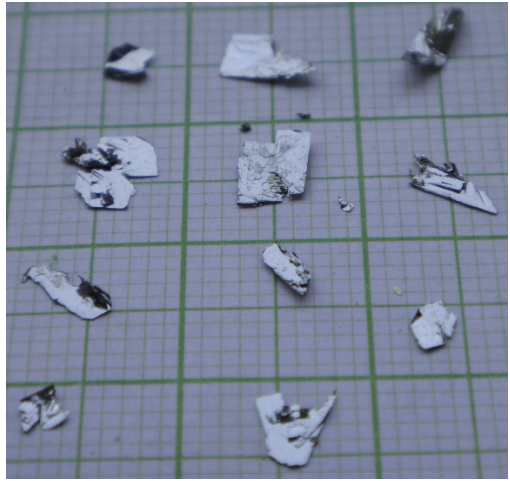

*Figure 2: Photograph of grown GeS$_{0.25}$Se$_{0.75}$ (I$_2$) crystals.*

**Table 1** Growth condition for GeS$_{0.25}$Se$_{0.75}$ (I$_2$) single crystals.

| Crystal | Temperature Distribution | | Growth Period (days) | Dimension of Grown Crystals (cm²) |
|---|---|---|---|---|
| | Reaction Zone (K) | Growth Zone (K) | | |
| GeS$_{0.25}$Se$_{0.75}$(I$_2$) | 883 | 843 | 7 | 0.8 × 0.9 |

**Primary Components of PEC Solar Cell**

Photoelectrochemical cells are solar cells that extract electrical energy from light, including visible light. Each cell consists of a semiconducting photo anode and a metal cathode immersed in an electrolyte. A typical PEC has three primary components; Semiconductor Electrode, Counter Electrode, and Electrolyte.

A glass rod of 0.5 cm in diameter and 10 to 12 cm in length with a narrow bore of diameter 0.05 cm was used to prepare the electrode. One end of this narrow-bore glass rod was flattened by a hot gas blow. The flat portion was used as a platform for resting the crystal. The narrow bore was used as a passage for traversing a good conducting copper wire. The copper wire was flattened at one end for getting in contact with the crystal. In the present work, a semiconductor electrode was fabricated in such a way that the contacting material (adhesive silver paste) provided good ohmic contact between the copper wire and the backside of the crystal. The whole assembly was then kept in an oven for a few hours at 100 ºC for baking. After proper setting of the crystal on the copper wire terminal, the semiconductor was covered with an epoxy resin (Araldite) leaving a light-exposed an area of 2-5 mm² for exposure to the light source. The so-prepared complete device semiconductor electrode is shown in Fig. 3.

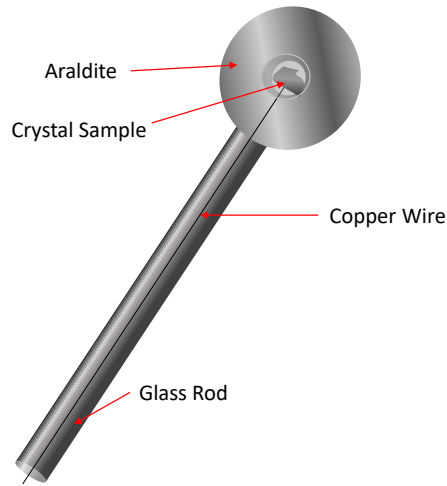

*Figure 3:* The semiconductor electrode

A counter electrode in PEC solar cells is required to complete the electrochemical reactions in a cell for better performance of the PEC solar cell. Generally, Platinum or graphite is a widely used material for the same. Many materials have been investigated electrochemically as counter electrode materials. A counter electrode in PEC solar cells is required to complete the electrochemical reactions in a cell for better performance of the PEC solar cell. Generally, Platinum or graphite is a widely used material for the same.

**Selection of Appropriate Electrolyte**

To obtain workable photoconversion from PEC solar cells, the selection of suitable electrolytes is very important. The electrolyte decides the band bending in the semiconductor near the interface and hence the efficiency of photoconversion. For this PEC investigation author have used electrolyte with the composition 0.025M $I_2$ + 0.5 M NaI + 0.5 M $Na_2SO_4$ the light intensity was kept at 30mW/cm² and gave the minimum dark voltage ($V_D$) and dark current ($I_D$) and as well provides the maximum value of photocurrent ($I_p$) and photovoltage. In this case, a mixture of iodine ($I_2$), sodium iodide (NaI), and sodium sulfate ($Na_2SO_4$) was employed as an electrolyte. All the chemical products were reagent grade and the electrolyte solutions were prepared using triple distilled water. The solutions were not stirred during the measurement. Here the Photoelectrode was prepared using $GeS_{0.25}Se_{0.75}$ ($I_2$) crystals having visibly smooth surfaces.

**Solar Cell Fabrication**

PEC solar cells with $GeS_{0.25}Se_{0.75}$ ($I_2$) photoelectrodes were fabricated in the same manner as described in [8]. The electrode was immersed in an appropriate electrolyte (to be discussed below) contained in a corning glass beaker. A platinum grid (3 cm x 3 cm) was used as a counter electrode. A schematic diagram showing the semiconductor electrode and PEC cell is given in Fig. 3 and Fig. 4 respectively. The cell was illuminated with light from an incandescent lamp. The intensity of illumination was measured using the light measuring instrument 'Luxmeter' (TES electrical electronic corporation TES 1332A). Photocurrent and photovoltage were recorded using digital multimeters. To vary the power point on V-I characteristics a series of variable resistance of different values have been used.

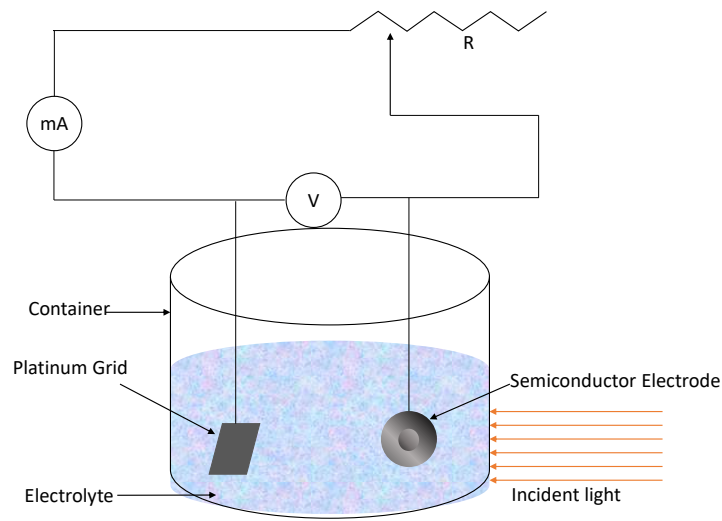

*Figure 4: Schematic diagram of PEC solar cell.*

## Results and Discussion

In order to see the effect of using electrodes with GeS$_{0.25}$Se$_{0.75}$ (I$_2$) single crystals on photoresponse in the PEC solar cell, the surface of all the electrodes was taken after cleaving them with an adhesive tape. Data for the photoresponse study of grown crystals used in PEC solar cells is given in Table 2. The current versus voltage characteristics drawn for the electrode is shown in Fig. 5. The Solar cell parameters e.g. I$_{sc}$, V$_{oc}$, the maximum power (P$_{max}$) the voltage at maximum power (V$_{max}$), the fill factor (F.F), and the efficiency ($\eta$) as determined from plots is shown in Table 3.

**Table 2:** Data for photoresponse study of grown crystals used in PEC solar cell

| **Particulars** | **GeS$_{0.25}$Se$_{0.75}$ (I$_2$)** |
| --- | --- |
| Type | P -type |
| Surface | Plane obtained after cleaving the as grown face |
| Contact | Silver Paste |
| Electrolyte | 0.025 M I$_2$ + 0.5 NaI + 0.5 M Na$_2$SO$_4$ |
| Area of the electrode's surface exposed | 3.14 ×10$^{-6}$ m$^2$ |
| Intensity | 30 mW/ cm$^2$ |

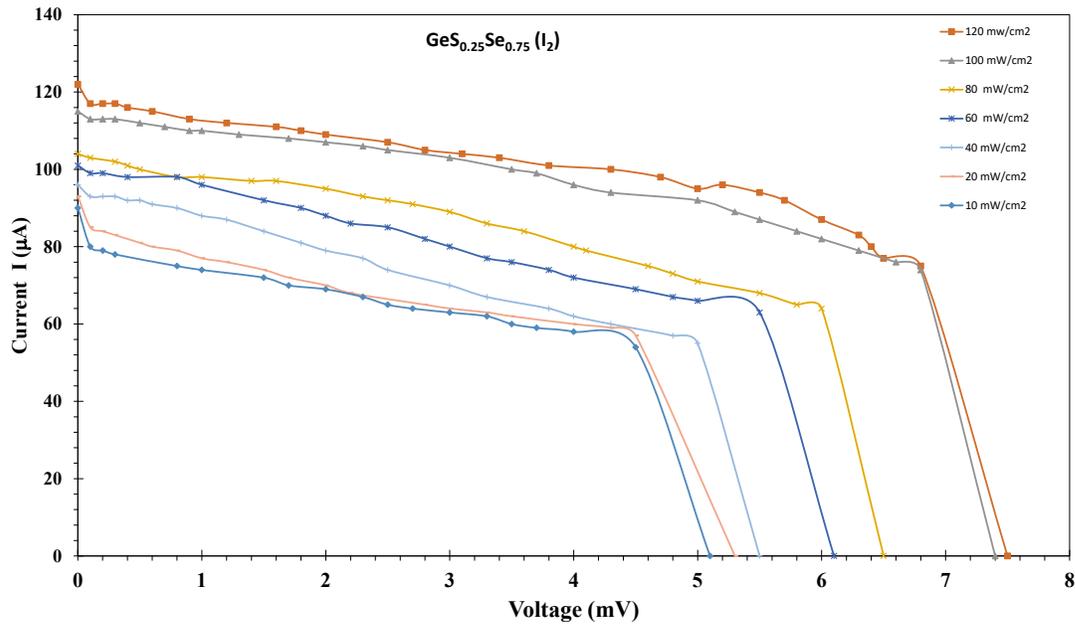

***Figure 5***: *I- V Characteristics for GeS$_{0.25}$Se$_{0.75}$ (I$_2$) single*

The photoresponse studies carried out in electrolytes of different compositions established that the electrolyte with the composition 0.025 M $I_2$ + 0.5 M NaI + 0.5 M $Na_2SO_4$ is the most suitable befitting electrolyte for the present work.

**Table 3:** Characteristic parameters of $GeS_{0.25}Se_{0.75}$ ($I_2$) based PEC solar cell with intensity illumination.

| Intensity (mW/cm²) | Short circuit current $I_{sc}$ (µA) | Open circuit voltage $V_{oc}$ (mV) | Power max. $P_{max}$ (µA mV) | Fill Factor (F.F) | Efficiency (η%) |
|---|---|---|---|---|---|
| 10 | 5.2 | 92 | 247.5 | 0.5173 | 0.505 |
| 20 | 5.3 | 93 | 256.5 | 0.5204 | 0.261 |
| 30 | 5.4 | 94 | 264.0 | 0.5201 | 0.179 |
| 40 | 5.5 | 96 | 275.0 | 0.5208 | 0.140 |
| 50 | 5.7 | 98 | 300.0 | 0.5371 | 0.122 |
| 60 | 5.8 | 100 | 318.6 | 0.5493 | 0.108 |
| 70 | 6.1 | 101 | 346.0 | 0.5616 | 0.100 |
| 80 | 6.3 | 102 | 365.4 | 0.4903 | 0.090 |
| 90 | 6.5 | 104 | 384.0 | 0.5976 | 0.087 |
| 100 | 6.9 | 108 | 442.0 | 0.6538 | 0.090 |
| 110 | 7.4 | 115 | 503.2 | 0.5913 | 0.093 |
| 120 | 7.5 | 122 | 510.0 | 0.5574 | 0.086 |

The variation of the short circuit current ($I_{sc}$) of fabricated PEC solar cells with the incident intensity of polychromatic light was investigated. It is quite clear from Table 3 that short circuit current ($I_{sc}$) Open Circuit Voltage ($V_{oc}$) and Maximum Power ($P_{max}$) increase with increasing intensity of the light. These observations are quite obvious because the short circuit current always depends upon many parameters like photogeneration of charge carriers within the semiconducting material and their effective separation, charge transfer process across the semiconductor electrolyte interface and the overall series resistance of a solar cell, etc. among all these parameters, the series resistance will affect the magnitude of short circuit current whereas the photogeneration of charge carriers and their effective separation and their contribution to the charge transfer mechanism decide the variation of short circuit current with illumination light. Fill factor is a parameter that judges the quality of PEC solar cell and photoconversion efficiency is also an important parameter of PEC solar cell.

As a matter of fact, PEC solar cells score over their solid-state counterparts on several points, some of them being ease of fabrication and in-built storage capacity. But, in order to make PEC solar cells viable, their conversion efficiencies have to be enhanced to reach optimum values. Several feasible efficiency enhancement processes for PEC solar cells such as electrode surface modification, photoetching, electrolyte modification, etc. are described and discussed by Pandey et al. and Patel et al. [9,10] have also pointed out that a high degree of perfection of the electrode surface is required to obtain higher solar energy conversion efficiency from a PEC solar cell. A method such as selection of electrode surfaces, use of proper ohmic contact, etching of electrode surfaces, use of optimum radiation for illuminating the cell, etc. has to be evolved to achieve the highest solar to electrical energy conversion from PEC solar cell fabricated with semiconducting crystals.

**Conclusion**

It is concluded from the work presented here that, the crystals of $GeS_{0.25}Se_{0.75}$ ($I_2$) were grown successfully with a chemical vapor transport technique using iodine as a transporting agent. The photoelectrochemical behavior was studied and it shows that with the increasing light intensity the solar cell parameters like short circuit current (Isc), open-circuit voltage (Voc), and maximum power (Pmax) also increase. From the data of fill factor (F.F.) and efficiency (n) it can be concluded that the sample yields less energy. Hence, to obtain an optimum value of photoconversion efficiency one has to evolve methods e.g. selection of electrode surface and its modification, etching of electrode surface, use of optimum radiation for illuminating the cell, etc.

**Acknowledgement**